\documentclass[12pt]{article}
\usepackage{graphics}
\begin{document}
\begin{center}
DYNAMICS OF PRIMORDIAL HYDROGEN RECOMBINATION WITH ALLOWANCE FOR A 
RECOIL FOR SCATTERING IN THE Ly-$\alpha$ LINE
\end{center}

\begin{center}
S.I.Grachev$^{1}$, V.K.Dubrovich$^{2,3,}$ 
\footnotemark[4]\footnotetext[4]{e-mail: dvk47@mail.ru}
\end{center}

\begin{center}
{\it
$^1$Sobolev Astronomical Institute, Saint Petersburg State University,
Saint Petersburg, Russia\\
$^2$Special Astrophysical Observatory of RAS, Saint Petersburg branch,
Saint Petersburg, Russia\\
$^3$The Main Astronomical Observatory of RAS, Saint Petersburg, Russia}\\
\medskip

Received/Accepted
\end{center}

\noindent {\it Abstract}. It is shown that taking into account a recoil
for radiation scattering in the Ly-$\alpha$ line can lead to a noticable
acceleration of primordial hydrogen recombination. Thus for 
$\Lambda$CDM model a decrease of ionization degree exceeds 1\% for redshifts z
in a range 800 -- 1050 achieving $\approx 1.3$\% at $z=900$. Corresponding
corrections to the cosmic microwave background power spectra can achieve 1.1\%
for $TT$ spectra and 1.7\% for $EE$ ones. Radiative 
transfer in these calculations was treated in a quasistationary approximation.
Numerical solutions are also obtained in diffusion approximation for a 
nonstationary problem
of Ly-$\alpha$ line radiative transfer under partial frequency redistribution
with a recoil. An evolution of a local line profile is traced to as well as an
evolution of a relative number of uncompensated transitions from 2p state down
to 1s one. It is shown that taking into account nonstationarity of Ly-$
\alpha$ line radiative transfer can lead to an additional acceleration
of primordial hydrogen recombination.
\medskip


\begin{center}
INTRODUCTION
\end{center}

In connexion with the proposed considerable increase of CMBR observations
precision the requirements grew as well to predictions accuracy of the theory
being used to interpret the observations. In particular this concerns the
theory of cosmological recombination. Required accuracy is about 0.1\%. So
at last time a few papers appear (Dubrovich, Grachev, 2005; Chluba, Sunyaev, 
2006; Kholupenko, Ivanchik, 2006; Rubi\~no-Martin et al., 2006; Chluba et al.,
2007; Wong, Scott, 2007; Chluba, Sunyaev, 2007) where recombination history
of primordial hydrogen is calculated with a proper allowance for some fine
effects which however are essential at the required level of accuracy.

In this paper we study an effect of allowance for a recoil under radiation
scattering in Ly-$\alpha$ line on recombination history of primordial hydrogen.
Earlier we already carried out calculations of such kind (Grachev, Dubrovich, 
1991) for purely baryonic models. An effect of a recoil turned out to be
practically negligible. However our calculations fulfiled in the framework
of present-day $\Lambda$CDM models show that for these models characterized by
low baryon density the effect of recoil becomes nearly ten times larger. 
The reason is in the fact that at low baryon density the relative role of 
Ly-$\alpha$ photons escape from the process of scattering due to Universe
expansion (so called "intrinsic escape") considerably increases as compared
with another main mechanism of irreversible recombination, namely,
two-photon transitions from the second level. Therefore the role of fine
effects in treating of radiation transfer in the Ly-$\alpha$ line grows 
correspondingly.

The second curcumstance to which we pay attention is an influence of
nonstationarity of radiaton transfer in the main resonance line on 
recombination history. Up to now a quasi-stationary approximation is in usage.
Namely, when computing hydrogen recombination history one makes use of 
stationary Sobolev approximation for the number of uncompensated transitions
from 2$p$ state down to 1$s$ one (so called net radiative bracket (NRB)).
However it would be very important to determine characteristic time necessary 
for NRB to reach a steady-state limit. In this connexion we fulfiled model 
numerical calculations of time dependent radiative transfer in a resonance
line under partial frequency redistribution (PFR) allowing for a recoil in the
process of scattering. It turned out that time of reaching a steady-state limit
for NRB may well exceed 1\% (in $\Delta z/z$ units) in the range of 
redshifts near $z\approx 1000$.

\begin{center}
RECOMBINATION DYNAMICS ASSUMING QUASI-STATIONARY RADIATIVE TRANSFER
\end{center}

It is known that Sobolev approximation widely used in astrophysics to treat
radiative transfer in spectral lines in moving media gives an exact solution
of a stationary problem for the case of two-lewel atoms in an infinite uniform
expanding medium assuming complete frequency redistribution
(CFR) in a single scattering. Then for a relative number of uncompensated
transitions (so called NRB) we have
\begin{equation}
r_{21}\equiv [n_2(A_{21}+B_{21}J_{12})-n_1B_{12}J_{12}]/n_2A_{21}=\beta (1-B/S),
\label{r21}
\end{equation} 
where  
\begin{equation}
\beta=\gamma\left(1-e^{-1/\gamma}\right)
\label{betaS}
\end{equation}
has a sense of photon escape probability out of the process of scattering,
\begin{equation}
\gamma=\frac{8\pi}{\lambda_{12}^3}\frac{H}
{A_{21}[(g_2/g_1)n_1-n_2]}
\label{gamma}
\end{equation}
is a dimensionless velocity gradient proportional to velocity gradient $H$
(which is the Hubble factor in cosmological context). A quantity $\tau_{\rm S}
=1/\gamma$ is called Sobolev optical distance. It is an optical depth of
an infinite expanding medium along photon path. Above we use standard 
designations for Einstein coefficients of transition probabilities and for
statistical wheights of levels. Moreover $J_{12}$ is the mean intensity of
radiation in a line, $\lambda_{12}$ is a transition wavelength, 
$S=(2hc/\lambda_{12}^3)[g_1n_2/(g_2n_1-g_1n_2)]$ is a line source function, 
$B$ is the Planck function at the line center (we suggest that far in the
violet wing of the line radiation intensity is planckian and we also ignore
frequency dependence of the Planck function within the line). From eqs. 
(\ref{betaS}) and (\ref{gamma}) we have equation $\beta\propto H/\Omega_{\rm B}$ 
which clearly demonstrates a grouth of escape probability when $\Omega_{\rm B}$ 
decreases.

As it was shown by Chugaj (1980) rejection of CFR i.e. usage of PFR
does not affect on a mean number of scatterings and on excitation degree of
atoms. Therefore if one neglects induced radiation and suggests that
$\gamma\ll 1$ then the equations given above may be used for PFR (Grachev, 1989;
Grachev, Dubrovich, 1991) if only recoil under scattering is neglected.
Allowing for recoil leads to the growth of photon escape probability due to
additional redshift of photons. From analytical solution of the problem 
in diffusion approximation (Grachev, 1989) the following expression is 
obtained for it suggesting zero boundary condition in a far violet wing:
\begin{equation}
\beta=i(-\infty)\gamma, \quad \gamma\ll 1,
\label{betar}
\end{equation}
where correction factor $i(-\infty)>1$. It depends on parameters
\begin{equation}
\rho=2\delta x_\gamma, \quad \sigma=x_\gamma/x_\lambda,
\label{rhos}
\end{equation}
where $\delta$ is a recoil parameter and $x_\gamma$ and $x_\lambda$ are
characteristic dimensionless frequencies:
\begin{equation}
\delta=h\nu_{12}/Mv_{\rm th}c,\quad x_\gamma=(3a\lambda/2\pi\gamma)^{1/3},
\quad x_\lambda=\sqrt{\lambda/2(1-\lambda)}. 
\label{xgam}
\end{equation}
Here $M$ is an atom mass, $v_{\rm th}=\sqrt{2kT_e/M}$ is a most probable 
velocity of atoms thermal motion, $a$ is the Voigt parameter, $\lambda$ is
an albedo per single scattering (a probability of photon "survival" in a
single scattering). Characteristic frequency $x_\gamma$ was introduced by
Chugaj (1980). It separates domain of photons diffusion ($|x|<x_\gamma$) from
domain of drift ($|x|>x_\gamma$) due to differential motion (expansion)
of the medium (here and so on $x=(\nu-\nu_{12})/
\Delta\nu_{\rm D}$, $\Delta\nu_{\rm D}=\nu_{12}v_{\rm th}/c$). Results of
numerical calculations of $i(-\infty)$ can be approximated by equations
\begin{equation}
i(-\infty)\approx 1+\rho/[1+(\sigma^2/3)(4-\sigma^2)/(2+\sigma^2)
+\sigma^4/6],\quad \sigma>0.38,
\label{infap}
\end{equation}
and
\begin{equation}
i(-\infty)\approx \left[3\int_0^\infty e^{-\rho y-y^3}y^2dy\right]^{-1}, \quad
\sigma<0.38,
\end{equation}
with an error less than 0.3\%. It should be noted that for $i(-\infty)-1$ an
error of these equations is less than 8\%.

For non-zero boundary condition $I(+\infty)=B$ in a far violet wing eq.
(\ref{r21}) turns then into (Grachev, Dubrovich, 1991)
\begin{equation}
r_{21}=\gamma [i(-\infty)-B/S].
\label{r21r}
\end{equation}
Computations of hydrogen recombination history using eqs. (\ref{infap}) --
(\ref{r21r}) for $2p$ -- $1s$ transition were carried out by us earlier in
the framework of 60-level model of hydrogen atom for purely baryonic universe
models and in particular for values of parameters being used in a pioneer
works on cosmological recombination (Zeldovich et al., 1968; Peebles, 1968): 
$\Omega_{\rm tot}=1$, $\Omega_{\rm B}=1$, $H_0=98$ km/s/Mpc, $T_0=2.7$ K, 
$Y_{\rm He}=0$. Results of these our calculations are shown in Fig. 1 along
with the results for $\Lambda$CDM model with parameters
$\Omega_{\rm tot}=1$, $\Omega_{\Lambda}=0.7$, $\Omega_{\rm DM}=0.26$,
$\Omega_{\rm B}=0.04$, $H_0=70$ km/s/Mpc, $T_0=2.728$ K, $Y_{\rm He}=0.24$.
Namely, in Fig. 1 there are graphs of relative variation (in percents) of
a fractional electron density $x_e=n_e/n_{\rm H}$ due to allowance for a recoil 
under Ly-$\alpha$ photons scattering. (Here $n_e$ and $n_{\rm H}$ are the number
densities of electrons and of hydrogen atoms and ions respectively.)
As a mechanism of Lyman photons "destruction"
per scattering we consider excitation and ionization of atoms from $2p$ state
due to absorption of background black-body radiation. So we have for albedo
per a single scattering an expression
\begin{equation}
\lambda = A_{2p,1s}\left/\left(A_{2p,1s}+R_{2p,c}+\sum_{i=3}^{\infty}R_{2p,i}
\right)\right..
\label{lambda}
\end{equation}
One can see in Fig. 1 that for a present-day Universe model recoil 
contribution increases more than ten times and is in the level of 1\% near
the surface of last scattering. 

Variation of primordial hydrogen recombination history has an influence on the
power spectra of CMBR angular fluctuations in intensity (temperature) and
polarization. Corresponding graphs are shown in Fig. 2. Computations were
fulfiled by means of CMBFAST package (Seljak, Zaldarriaga, 1996). It is seen
that relative variations can reach 1.1\% for $TT$ spectra and 1.7\% for $EE$
spectra for multipoles $l<2500$.

\begin{figure}[t]

\vspace*{-6cm}
\centering

\resizebox{1.0\textwidth}{!}{\includegraphics{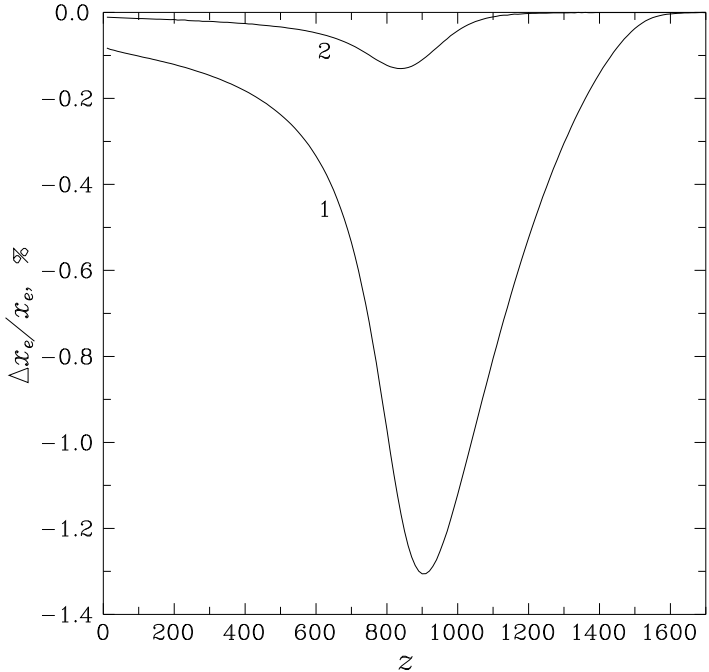}}

\vspace{-10cm}

\caption{Variation of hydrogen recombination history due to effect of a recoil
under scattering in Ly-$\alpha$ line. Parameters values: 1) $\Omega_{\rm tot}=
1$, $\Omega_{\Lambda}=0.7$, $\Omega_{\rm B}=0.04$, $\Omega_{\rm DM}=0.26$, 
$H_0=70$ km/s/Mpc, $T_0=2.728$ K, $Y_{\rm He}=0.24$; 2) $\Omega_{\rm tot}=1$, 
$\Omega_{\Lambda}=0$, $\Omega_{\rm B}=1$, $\Omega_{\rm DM}=0$, $H_0=98$ 
km/s/Mpc, $T_0=2.7$ K, $Y_{\rm He}=0$.}
\end{figure}

\begin{figure}[t]

\vspace*{-6cm}
\centering

\resizebox{1.0\textwidth}{!}{\includegraphics{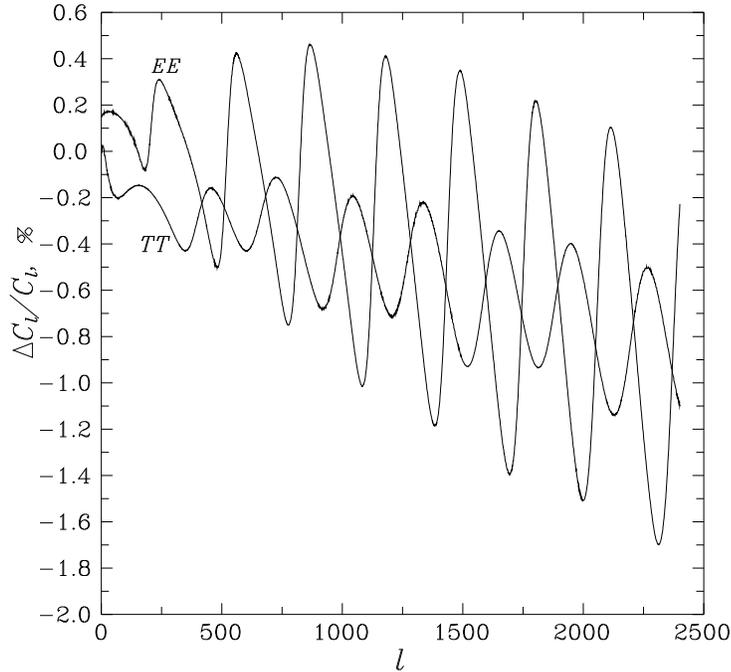}}

\vspace{-10cm}

\caption{Variation of CMBR power spectra of temperature ($TT$) and 
polarization ($EE$) fluctuations. Computations are fulfiled with the
CMBFAST package (Seljak, Zaldarriaga, 1996). Values of parameters are the same 
as for the curve 1 in Fig. 1.}

\end{figure}

\begin{center}
NONSTATIONARY RADIATIVE TRANSFER in Ly-$\alpha$ LINE
\end{center}


To clear up how fast NRB reachs a steady-state limit we have solved a model
problem of nonstationary radiative transfer in isolated spectral line. It is
suggested that at some initial moment of time ($t=0$) there is no radiation
and at this moment primary sources of line radiation with a constant power are
switched on. The main integro-differential equation of radiative transfer has
a form
\begin{equation}
\frac{\partial I(t,x)}{\partial t}-\gamma
\frac{\partial I(t,x)}{\partial x}=-
\phi(x)I(t,x)+\lambda\int_{-\infty}^{+\infty}R(x,x')
I(t,x')dx' +(1-\lambda)\phi(x)S_0,
\label{mtreq}
\end{equation}
where $I(t,x)$ is a radiation intensity ($x=(\nu-\nu_{12})/\Delta\nu_{\rm D}$ 
is a distance from a line center measured in Doppler widths; $t$ is a time
measured in units of mean time between succesive scatterings), $R(x,x')$ is
a frequency redistribution function per scattering, $\phi(x)$ is an absorption
coefficient profile, $\gamma$ is a dimensionless velocity gradient (see
above eq. (\ref{gamma}) in which however a contribution of induced radiation  
should be neglected), $\lambda$ is an albedo per a single scattering. The 
following normalizations hold
\begin{equation}
\int_{-\infty}^{+\infty}R(x,x')dx=\phi(x'),\quad \int_{-\infty}^{+\infty}
\phi(x)dx=1.
\label{normR}
\end{equation}
Initial and boundary conditions for eq. (\ref{mtreq}) are $I(0,x)=0$,
$I(\tau,+\infty)=0$.

In the context of cosmological recombination an absorption coefficient profile
is the Voigt one with parameter $a=A_{2p,1s}/4\pi\Delta\nu_{\rm D}$. As 
concerned frequency redistribution function with allowance for a recoil it
was derived in the paper by Basko (1981) where diffusion approximation was
also obtained for the integral term (in transfer equation) describing an act
of a single scattering. Using this approximation one can reduce eq. 
(\ref{mtreq}) (with $S_0=1$)  to the form
\begin{equation}
\frac{\partial I}{\partial\tau}-\frac{\partial I}{\partial y}=
-\left(\sigma^2+\frac{2\rho}{y}\right)\frac{I(\tau,y)}{3y^2}+
\frac{1}{3y^2}\left(\rho-\frac{2}{y}\right)\frac{\partial I}{\partial y}+
\frac{1}{3y^2}\frac{\partial^2 I}{\partial y^2}+\frac{\sigma^2}{3y^2}+
\frac{\sigma^2}{3\varepsilon}[1-I(\tau,0)]\delta(y),
\label{difeq}
\end{equation}
where new dimensionless frequency and time are $y=x/x_{\gamma}$, $\tau=\gamma 
t/x_{\gamma}$ and $\varepsilon=a/\pi x_{\gamma}$ is a small parameter. 
Parameters $\rho$ and $\sigma$ are defined above by eqs. (\ref{rhos}) and 
(\ref{xgam}). When obtaining eq. (\ref{difeq}) the profile of absorption
coefficient was taken approximately in the form (Grachev, 1989)
\begin{equation}
\phi(x)=\delta(x)+a/\pi x^2,
\label{phi}
\end{equation}
where the first term ($\delta$-function) describes Doppler core and the second
one corresponds to the Lorentz wings of the Voigt profile.

In a steady-state case (when $\partial I/\partial\tau=0$) analytical solution
of eq. (\ref{difeq}) was constructed by Grachev (1989) and then has been used
(Grachev, Dubrovich, 1991) to compute recombination history of primordial
hydrogen (see also above in the preceeding section). In the present paper
we obtain numerical solutions of nonstationary eq. (\ref{difeq}) for the
values of parameters typical for cosmological recombination. (As a matter of
fact the function $f(\tau,y)=\int_y^\infty I(\tau,y')dy'$ was found at first 
from equation which is easily obtained by integrating eq. (\ref{difeq}).)
We make use of numerical method of solving nonstationary problems of
radiative transfer theory proposed by Grachev (2001). Results are in Fig. 3. 
The values of $\Lambda$CDM model parameters are the same as for the curve
1 in Fig. 1. As for parameters depending on redshift $z$ their values were
calculated by means of our code which computes hydrogen recombination history
(Grachev, Dubrovich, 1991). The point $z=1180$ near the surface of last
scattering was taken. At this point we have $x_e=0.300$, $1-\lambda=2.00\cdot 
10^{-4}$, $a=8.30\cdot
10^{-4}$, $\gamma=2.26\cdot 10^{-9}$, $x_{\gamma}=56.0$, $\varepsilon=5\cdot
10^{-6}$, $\rho=0.0500$, $\sigma=1.12$, $i(-\infty)-1=0.0296$. For comparison
in Fig. 4 results of similar calculations are shown but for $\sigma=0$ which 
corresponds to conservative scattering ($\lambda=1$) (then $i(-\infty)-1=
0.0455$). (Since in this case eq. (\ref{difeq}) becomes homogeneous we 
normalize its solution to unity at $\tau=\infty$ and $x/x_{\gamma}=-\infty$.) 
One can see that in both cases the profiles approach asymptotically to a steady
state limit which corresponds to the upper curves on Figs. 3 and 4.

\begin{figure}[t]

\vspace*{-6cm}
\centering

\resizebox{1.0\textwidth}{!}{\includegraphics{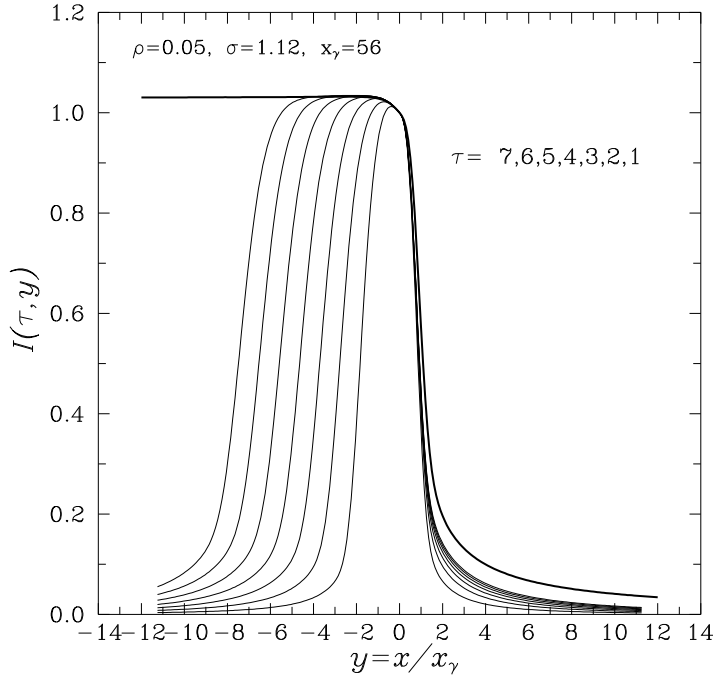}}

\vspace{-10cm}

\caption{Evolution of a local Ly-$\alpha$ line profile in the case of 
non-conservative scattering. An upper profile corresponds to $\tau=\infty$. 
Values of model parameters are the same as for the curve 1 in Fig. 1.} 
\end{figure}
\begin{figure}[t]

\vspace*{-6cm}
\centering

\resizebox{1.0\textwidth}{!}{\includegraphics{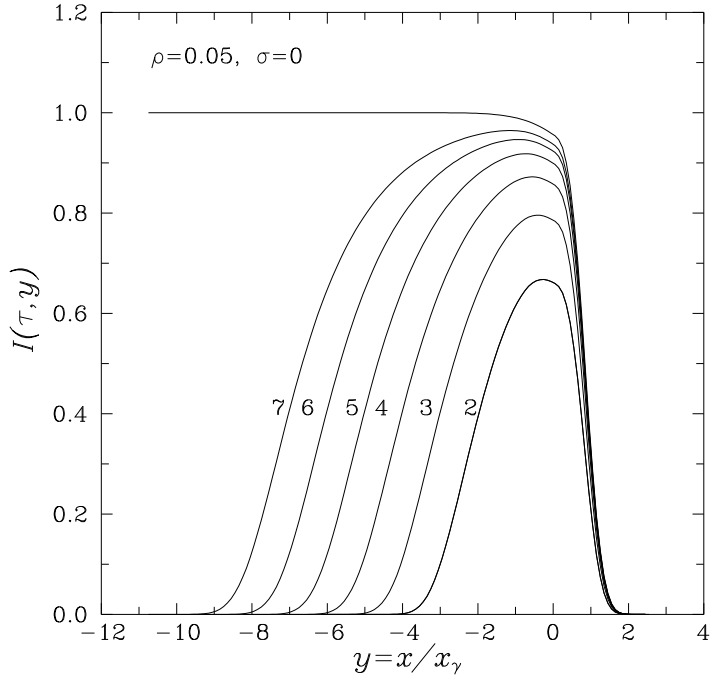}}

\vspace{-10cm}

\caption{The same as in the preceeding Fig. but for conservative scattering. 
Numbers near the curves are the moments of time $\tau$.} 
\end{figure}

For NRB one can obtain directly from eq. (\ref{mtreq}) the following 
expression
\begin{equation}
r_{21}=\beta=\frac{\gamma}{1-\lambda+\lambda J(\tau)}\frac{\partial}{\partial
\tau}\int_{-\infty}^{+\infty}I(\tau,y)dy,
\end{equation}
where
\begin{equation}
J(\tau)=\int_{-\infty}^{+\infty}\phi(x)I(\tau,x)dx.
\end{equation}
According to eq. (\ref{phi}) it follows from here that $J(\tau)=I(\tau,0) + 
O(\varepsilon)$. Fig. 5 displays that approaching of $r_{21}$ to the 
stationary limit (dashed horizontal lines) is going on quite slowly especially
for non-conservative scattering ($\sigma=1.12$). The point is that for
non-conservative scattering the line profile has extended quasi-lorentzian
wings (one can show that $I(\tau,y)\sim 
\sigma^2\tau [3y(y+\tau)]^{-1}$ for $|y|\gg 1$, $|y+\tau|\gg 1$) which evolve
very slowly towards the steady state limit ($I(\infty,y)\sim
\sigma^2/3y$) for $y\gg 1$ (see Fig. 3). According to Fig. 5 the time to reach
steady-state limit can be as large as $\tau\approx 10$ and even more. An 
interval of redshifts $\Delta z$ corresponding to this time can be estimated
from equation $dz/dt=-(1+z)H(z)$ from which it follows that
\begin{equation}
\frac{\Delta z}{1+z}=\frac{v_{\rm th}(z)}{c}x_{\gamma}(z)\Delta\tau.
\label{deltz}
\end{equation}
From here one can obtain that $\Delta z/z=1.4\cdot 10^{-3}\Delta\tau$ for 
given above values of parameters at $z=1180$. Thus non-stationarity of 
radiative transfer in Ly-$\alpha$ line can lead to an appreciable acceleration 
of irreversible recombination of primordial hydrogen.

From physical point of view an increase of NRB due to non-steady state 
radiative transfer takes place because the line does not reach a saturation
in some frequency regions after primary sources switch on, i.e. the are no
enough resonance photons in these regions to compensate in a maximum degree
radiative transitions down by those of up.

\begin{figure}[t]

\vspace*{-6cm}
\centering

\resizebox{1.0\textwidth}{!}{\includegraphics{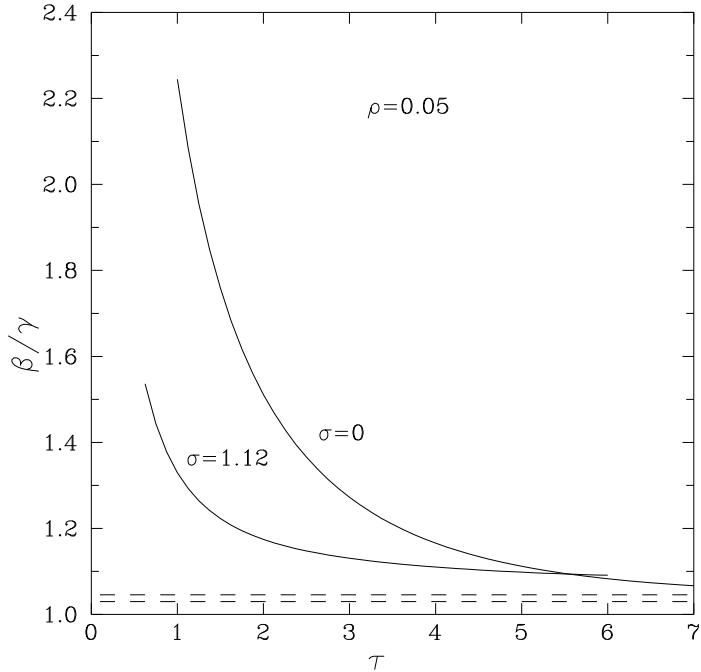}}

\vspace{-10cm}

\caption{Ly-$\alpha$ photons escape probability (or NRB) normalized to Sobolev
value. Dashed horizontal lines correspond to the steady-state limit allowing
for a recoil (upper line: $\sigma=0$, lower line: $\sigma=1.12$). }
\end{figure}

\begin{center}
CONCLUSION
\end{center}

In this paper we study an effect of a more accurate treating of radiative 
transfer in Ly-$\alpha$ line on a calculated recombination history of 
primordial hydrogen. An urgent necessity of such kind investigation is
connected with increased role of Lyman photons escape from a 
scattering process in the framework of present day $\Lambda$CDM models
characterized by low baryon density. We show that alowance for a recoil under
scattering in Ly-$\alpha$ line can lead to an appreciable acceleration
of primordial hydrogen acceleration. As a main one we consider a model with
the following values of parameters $\Omega_{\rm tot}=1$, $\Omega_{\Lambda}=
0.7$, $\Omega_{\rm DM}=0.26$, $\Omega_{\rm B}=0.04$, $H_0=70$ km/s/Mpc, 
$T_0=2.728$ K, $Y_{\rm He}=0.24$. For this model a decrease of hydrogen 
ionization degree exceeds 1\% in the range of redshifts $z=800 - 1050$
reaching $\approx 1.3$\% at $z=900$. Corresponding variations in computed
power spectra of CMBR angular fluctuations reach 1.1\% for $TT$ spectra and
1.7\% for $EE$ spectra for multipoles $l<2500$. It should be stressed that
radiative transfer in these calculations is treated (as well as in other
works on cosmological recombination of hydrogen) in a quasi-stationary
approximation. In order to study an applicability of this approximation
we have obtained numerical solutions of nonstationary problem of radiative
transfer in Ly-$\alpha$ line assuming partial (true) frequency redistribution 
with a proper allowance for a recoil. An evolution of a local line profile
is traced through as well as an evolution of number of uncompensated 
transitions from $2p$
state down to $1s$ one. It is shown that taking into acount non-stationarity
of radiative transfer in Ly-$\alpha$ line can lead to an apreciable 
additional acceleration of primordial hydrogen recombination.

{\it Acknowledgements}. The authors thank V.G.Romanyuk for computations of CMBR 
power spectra. We are also grateful to anonymous referee for his remarks on the
manuscript. This work is supported by RFBR grants 05-02-16658 and
06-02-17096 and by grant NSh-8542.2006.2 of the President of Russian 
Federation for Leading Scientific Schools support.

\begin{center}
REFERENCES
\end{center}

\begin{enumerate}
\item Basko M.M., Astrophysics, {\bf 17}, 69 (1981).
\item Chluba J., Rubi\~no-Martin J.A., Sunyaev R.A., MNRAS, {\bf 374}, 1310 
      (2007).
\item Chluba J., Sunyaev R.A., Astron. Astrophys., {\bf 446}, 39 (2006).
\item Chluba J., Sunyaev R.A., arXiv: astro-ph/0702531v2 
      (2007).
\item Chugaj N.N., PAZh., {\bf 6}, 166 (1980).
\item Dubrovich V.K., Grachev S.I., Astronomy Letters, {\bf 31}, 359 (2005).
\item Grachev S.I., Astrophysics, {\bf 30}, 211 (1989).
\item Grachev S.I., Astrophysics, {\bf 44}, 505 (2001).
\item Grachev S.I., Dubrovich V.K., Astrophysics, {\bf 34}, 124 (1991).
\item Kholupenko E.E., Ivanchik A.V., Astronomy Letters, {\bf 32}, 795 (2006).
\item Peebles P.J., Astrophys. J., {\bf 153}, 1 (1968).
\item Rubi\~no-Martin J.A., Chluba J., Sunyaev R.A., MNRAS, {\bf 371}, 1939 
      (2006).
\item Seljak U., Zaldarriaga M., Astrophys. J., {\bf 469}, 437 (1996).
\item Wong W.Y., Scott D., MNRAS, {\bf 375}, 1441 (2007).
\item Zel'dovich Ya.B., Kurt V.G., Sunyaev R.A., Zh. Eksp. Teor. Fiz., {\bf 55}, 
      278 (1968); English translation, Soviet Phys.--JETP Lett., {\bf 18}, 146
      (1969).

\end{enumerate}

\end{document}